\documentclass[aps,prb,twocolumn,floats,showpacs,superscriptaddress]{revtex4-1}

\usepackage{graphicx,epsfig}
\usepackage{times}
\usepackage{graphics,dcolumn,bm,float}
\usepackage{amssymb,amsmath,rotate,color}
\usepackage[title,titletoc,toc]{appendix}
\usepackage{graphicx}
\usepackage{wrapfig}

\usepackage{lipsum}
\usepackage{mwe}

\usepackage[pagebackref=false,colorlinks,linkcolor=blue,citecolor=blue,urlcolor=magenta]{hyperref}

\DeclareMathOperator\sgn{sgn}

\usepackage[mathlines]{lineno}
\usepackage{hyperref}
\usepackage{breqn}
\usepackage{bbold}
\usepackage{pgfplots}
\usepackage{float}
\usepackage{tkz-euclide}
\usepackage{braket}
\usepackage{physics}
\usepackage[caption=false]{subfig}
\usepackage[export]{adjustbox}
\usepackage{tikz}
\usetikzlibrary{through,calc}
\usetikzlibrary{positioning}

\newcommand{\om}{\Omega}
\newcommand{\be}{\begin{equation}}
\newcommand{\ee}{\end{equation}}
\newcommand{\ep}{\epsilon}

\newcommand{\bearr}{\begin{eqnarray}}
\newcommand{\eearr}{\end{eqnarray}}
\newcommand{\nn}{\nonumber}

\newcommand{\bsq}{{\boldsymbol{q}}}
\newcommand{\bsk}{{\boldsymbol{k}}}

\newcommand{\bs}{\boldsymbol}

\begin{document}
\preprint{}
\title{Tilt-induced kink in the plasmon dispersion of two-dimensional Dirac electrons}
\author{Z. Jalali-Mola}
\email{jalali@physics.sharif.edu}
\affiliation{
Department of Physics$,$ Sharif University of  Technology$,$ Tehran 11155-9161$,$ Iran
}

\author{S.A. Jafari}
\email{jafari@physics.sharif.edu}
\affiliation{
Department of Physics$,$ Sharif University of  Technology$,$ Tehran 11155-9161$,$ Iran
}
\affiliation{School of Physics$,$ Institute for Research in Fundamental Sciences (IPM)$,$ Tehran 19395-5531$,$ Iran}

\date{\today}

\begin{abstract}
The list of two dimensional Dirac systems with a tilt in their Dirac cone spectrum is expanding,
and now in addition to organic system $\alpha$(BEDT-TTF)$_2$I$_3$ includes the two dimensional $8Pmmn$-borophene sheet,
which allows for controlled doping by the gate voltage. 
We analytically calculate the polarization function of tilted Dirac cone for arbitrary
tilt parameter, $0\le \eta<1$, and arbitrary doping. This enables us to find two interesting plasmonic
effects solely caused by the tilt: (i) In addition to the standard plasmon oscillations, 
strong enough tilt, induces an additional linearly dispersing overdamped branch of plasmons, which is strongly Landau damped
due to overlap with a large density of intra-band free particle-hole (PH) excitations. (ii) There appears 
a kink in the plasmon dispersion for any non-zero tilt parameter. The kink appears when the plasmon branch enters the 
inter-band continuum of PH excitations. This kink becomes most 
manifest for wave vectors perpendicular to the tilt direction and fades away by approaching the tilt direction.
Experimental measurement of the wave vector and energy of the plasmonic kink, when combined with our analytic formula for the
kink energy scale, allows for a direct experimental measurement of the tilt parameter. 
\end{abstract}



\maketitle
\narrowtext

\section{introduction}

Dirac materials are now quite ubiquitous in condensed matter physics~\cite{WehlingReview},
and include one~\cite{Fradkin}, two~\cite{Novoselov2004} and three spatial dimensions~\cite{WehlingReview,Armitage2018}. 
In two dimensions a well known example of 
Dirac material is graphene~\cite{Bostwick2007,castro,Adam2011}, the two spatial dimensional character of which
allows for functionalization~\cite{Georgakilas} and manipulations. 
The interesting thing about condensed matter Dirac systems is that -- unlike the high energy physics --
it can be deformed in many ways. The deformation can induce anisotropy in the velocity $v_x,v_y$~\cite{Son2010},
or can deform the Brillouin zone, which then moves the Dirac cones in the Brillouin zone~\cite{Si2016}. 
One interesting deformation of the Dirac cone is to tilt it. In the case of graphene, a very small
amount of tilting can be obtained through coupling to lattice deformations~\cite{Cabra2013}. 
The first example of tilt in the Dirac cone  was realized
in the molecular organic material $\alpha$-(BEDT-TTF)$_2$I$_3$~\cite{Suzumura2006,Tajima2006}.
The smaller velocity scales of molecular orbitals compared to atomic $p_z$ orbitals of graphene
makes the tilt perturbation quite large in these systems~\cite{Suzumura2006,Suzumura2014,Goerbig2008}.

The layered organic conductor $\alpha$-(BEDT-TTF)$_2$I$_3$,  is one of 2D Dirac cone materials which consists of conducting layers of BEDT-TTF molecules separated by insulating layers of $I^-_3$
anions which conduction electrons are arranged on the square lattice with four molecules BEDT-TTF per unit cell~\cite{Suzumura2014}.
As the layers are weakly coupled to each other, the material  under the hydrostatic pressure above $1.5$ GPa is a
quasi-two-dimensional zero gap conductor with linear dispersion.
However, the dominant difference between the characteristic of graphene and organic conductor is that the graphene dispersion is rotationally invariant along the axis of the cone but the organic conductor is not rotationally invariant along the tilt axis. Moreover, the tilt axis is not normal to the $\bsk$ plane. 
 In addition, unlike graphene which its two Dirac cones are located at K, K', in organic conductor, Dirac cones have been located in general points $\pm\bsk_0$. 
Electronic structure calculations suggest anisotropic velocity show that the $v_x , v_y\simeq 10^5$~\cite{Katayama9, Suzumura2014,Goerbig2014,Kobayashi2009}. Empirically
determined velocities are slightly less than these values~\cite{Suzumura2014}. 
The tilted anisotropic Dirac equation that describes the low energy band structure in this system, 
has been confirmed by ab initio calculations\cite{Suzumura2014,Suzumura2006,Kino2006,Kobayashi2007,Ishibashi2006,Alemany2012}. 

The tilting properties can be induced in graphene by lattice deformation, that is accompanied by rotation symmetry breaking~\cite{Cabra2013,Mao}, 
which is more difficult and gives rise to a small tilt. However, elemental boron -- just to the left of Carbon in the periodic table -- can also afford to form a 
two-dimensional allotrope~\cite{Zhou2014}. The stable structure of borophene is the so-called orthorhombic $8-Pmmn$. 
This structure  with two non-equivalent buckled sub-lattices~\cite{Lopez2016}, and possesses tilted anisotropic 
Dirac cone which has been proven by {\em ab-initio} calculation~\cite{Lopez2016,Feng2017}. 
The borophene is now synthesized on Ag(111) surface and features anisotropic Dirac cone~\cite{Mannix15}. 
In this material the velocities $(v_x ,v_y) \simeq (0.86,0.69)\times 10^6 m/s$~\cite{Zabolotskiy2016}. 
Therefore the kinetic energy scale of Dirac electrons in borophene is less than graphene,
while in the organic materials they are at least an order of magnitude smaller than the kinetic energy 
of Dirac electrons in graphene. 
This already signals that the many-body and fluctuations phenomena in tilted Dirac cone materials must be even more 
profound than graphene~\cite{Fritz2017tilt}. 

The two-dimensionality of the latest $8Pmmn$-borophene allows for controlled
doping by the gate voltage. Therefore this is time to look into collective excitations of this system. 
For this one needs an analytical understanding of the polarization function that describes the density-density
correlations in the system. This fundamental quantity has been thoroughly calculated for the up-right Dirac
cone in graphene~\cite{hwang2007,Wunsch2006}. In the case of tilted Dirac cone, Nishine and coworkers have
given the analytical formula for the imaginary part of the polarization function~\cite{Nishine2010}. The real part
in their work is numerically calculated via the Kramers-Kronig relation from the imaginary part~\cite{Nishine2010}.
In a recent work, Sadhukhan and Agrawal have attempted the analytical calculation of the polarization function.
However, the determination of the signs and Fermi step functions in their work has not been correctly accomplished~\cite{Agrawal}. 
The above two works do not agree with each other. 
In this work, as will be detailed in the appendix, we carefully obtain analytical representation of both real and imaginary
parts of the polarization function for tilted Dirac cone at arbitrary doping. Our results agree with the numerical results 
of Ref.~\onlinecite{Nishine2010} for the real part. 

Our analytical result allows for detailed study of the plasmon excitations in tilted Dirac cone system. 
First of all, we find that the when the standard plasmon branch enters the inter-band portion of the PH continuum (PHC),
develops a kink. Again our analytic formula allows showing that the Landau damping in the inter-band PHC is negligible.
Therefore the plasmon branch on both sides of the kink will be long-lived, and can be experimentally determined. 
The entire plasmon structure is anisotropic, and the 
kink is most manifest for wave vectors perpendicular to the tilt direction. We suggest that a knowledge of the
wave vector and frequency of the kink in the hindsight allows for direct determination of the tilt parameter
from angular resolved electron energy loss spectroscopy (EELS). 
Our analytic formula further allows us to find that for large enough tilt parameters, there is another
branch of plasmon excitations inside the intra-band PHC which is overdamped due to a very large density of states (DOS) of
PH excitations. 

This paper has organized as follows: In section II we formalize the tilt and derive our
analytic representation of the polarization function. In Section III, we give a qualitative
and discussion of the role of tilt in plasmonic properties. In section IV, we identify the kink
in the plasmon dispersion and explain the physics behind it and suggest it as a way to 
experimentally measure the tilt. In section V, we provide asymptotic formula to address
plasmons and static screening in presence of kink. Appendices give very details of the calculation
to enable the reader to re-derive our results. We end the paper with the summary of findings.

\section{tilted Dirac cone model}
Effective theory of of massless tilted Dirac fermions is given by the following deformation of the
Dirac equation~\cite{Nishine2010,Goerbig2008},
  \begin{equation}
  H(k)=\hbar \begin{pmatrix}  v_{x0}k_x + v_{y0}k_y &  v_x k_x-i v_y k_y\\    v_x k_x+i v_y k_y &  v_{x0}k_x + v_{y0}k_y \end{pmatrix},
    \label{matrixform}
    \end{equation}
 where the off-diagonal (Fermi) velocities $v_x,v_y$, if different, stand for anisotropy, and diagonal velocities, $v_{x0}, v_{y0}$ 
represent the tilting characteristic of system. 
If we consider $v_{x0}=0$ and $v_{y0}=0$, the isotropic limit with $v_x=v_y=v_F$, will reduce 
this model to the graphene Hamiltonian. Through following transformation, 
\bearr
 &&\tilde{k}_x=k_x \cos \theta_{t} + \frac{k_y}{\gamma^2} \sin \theta_{t}, \nn\\&&
  \tilde{k}_y=-k_x \sin \theta_{t} + \frac{k_y}{\gamma^2} \cos \theta_{t},
  \label{transformation}
 \eearr
the tilted Dirac cone Hamiltonian Eq.~\eqref{matrixform}, can be rewritten as,
 \be
 H(\tilde{k})= \hbar v_x \begin{pmatrix}  \eta \tilde{k}_x &  \tilde{k}_x-i\tilde{k}_y\\   \tilde{k}_x+i\tilde{k}_y &  \eta \tilde{k}_x  \end{pmatrix}=\hbar v_x (\eta \tilde{k}_x \sigma_0+\tilde{\bsk}.\bs{\sigma}).
     \label{nmatrixform}
 \ee 
In Eq.~\eqref{transformation}, the dimensionless parameter $\eta$ and $\theta_t$ determine the tilting characteristic of system and defined as,
 \be
 \eta=\sqrt{\frac{v_{x0}^2}{v_x^2}+\frac{v_{y0}^2}{v_y^2}} ,~~~~\gamma=\sqrt{v_x/v_y},~~~~
 \cos \theta_t=\frac{v_{x0} }{v_x \eta}.
 \ee
Here $\gamma$ is intrinsic anisotropy, and the tilt parameter is given by $0\leq\eta\leq1$. The $\eta=0$ and $\gamma=1$ 
corresponds to the graphene~\cite{Katayama2008,Suzumura2014}.  

 The eigenvalues and eigenstates of  the transformed Hamiltonian are given by,
\be
E_{\lambda}(\tilde{k})= \hbar v_x \tilde{k} (\lambda+\eta \cos \tilde{\theta}_{\tilde{k}}) ~~~,~~~
 \ket{\bs{\tilde{k}},\pm}= \begin{pmatrix} 1 \\ \pm e^{i\tilde{\theta}_{\tilde{k}}} \end{pmatrix},
 \label{dispersion.eqn}
 \ee
where $\lambda=\pm$ refers to positive ($E_+$) and negative ($E_-$) energy branches, and 
$\tilde{\theta}_{\tilde{k}} $ is polar angle of the wave vector, $\tilde{\bsk}$, with respect to the $x$ axis. 
Note that the angular dependence in Eq.~\eqref{dispersion.eqn} persists even when the anisotropy generated by non-equal $v_x,v_y$
is not present (i.e. when $\gamma=1$). It is the genuine anisotropy due to tilting, as it vanishes when $\eta$ does. 
In what follows, to avoid cluttering up with notation, we replace the notation $\tilde{\bsk}$ with $\bsk$,
and similarly $\tilde\theta_{\tilde k}$ with $\theta_k$. 

The polarization function in linear response theory is defined by,
  \be
     \chi(\bs{q},\omega)=\int\frac{dt}{2\pi i}e^{i\omega t} \Theta(t)\langle [\rho_{\bs q}(t), \rho_{-\bs q}(0)] \rangle,
  \ee
the Lehman representation of which is given by,  
 \bearr
&& \chi(\boldsymbol{q},\omega)=\\&&   \frac{g  \gamma^2}{A v_x^2}  \lim\limits_{\ep\rightarrow 0}  \sum_{k,\lambda , \lambda'=\pm} \frac{n_{k,\lambda}-n_{k+q,\lambda'}}{\hbar \omega+E_{k,\lambda}-E_{k+q,\lambda'}+i\ep} f_{\lambda, \lambda'} (\bsk, \bsk')\nn.
\label{pai}
 \eearr
Here, matrix element of density operator between two eigenstate of $|\bsk,\lambda\rangle$ and $|\bsk',\lambda'\rangle$ is defined by the 
form factor $f_{\lambda, \lambda'} (\bsk, \bsk')$. The spin and valley degeneracy which are equal  to $2$, are included in constant coefficient $g$. The factor 
of $\gamma^2$ comes from the Jacobian of the transformation, Eq.~\eqref{transformation}.  $A$  is area of the two-dimensional system, and $\ep$ is 
defined as an infinitesimal positive constant. The Fermi distribution function is denoted by  $n_{k,\lambda}$, which at zero temperature 
reduces to step function. The wave vectors are related by $\bsk'=\bsk+\bsq$, with $\bsq$ being the momentum transfer, the direction of which
with respect to $x$ axis is $\phi$.
In the following, we analytically calculate this polarization function and upgrade a numeric calculation of an existing calculation~\cite{Nishine2010}
to an analytical expression which is benchmarked against the numerical calculation of Ref.~\onlinecite{Nishine2010}. Our results for the imaginary part
of the polarization function is identical to that in Ref.~\onlinecite{Nishine2010}. While the authors of this reference use Kramers-Kronig
relation to numerically calculate the real part of the polarization function, we are able to find analytic expressions for the real part, which
agrees with the numerical calculations of Ref.~\onlinecite{Nishine2010}. But our result does not agree with a recent calculation~\cite{Agrawal}.

\subsection{Undoped tilted Dirac cone}
In the undoped tilted Dirac cone, which corresponds to ($\mu=0$), the states with negative (positive) energy, which are in the 
lower (upper) part of the cone are always occupied (unoccupied). Hence the Fermi distribution function at zero temperature will 
be one (zero) for the valence (conduction) band states. Therefore, the polarization function Eq.\eqref{pai} reduces to, 
 \bearr
\chi_{0}(\boldsymbol{q},\omega)=&&\frac{g \gamma^2}{A} \lim\limits_{\ep\rightarrow 0} \sum_{k} f_{+,-}\left(\boldsymbol{k},\boldsymbol{q}\right)\times \nn\\&&\bigg\{ \frac{1}{\hbar \omega+E_{\bsk,-}-E_{\bsk+\bsq,+}+i\ep}  \nn\\&&-\frac{1}{\hbar \omega+E_{-\bsk-\bsq,+}-E_{-\bsk,-}+i\ep} \bigg\}.
 \label{paiud.eqn}
 \eearr
Here subscribe $0$ in $\chi_0$ stands for undoped tilted Dirac cone, and $f_{-,+}$ is the inter-band form factor, 
\bearr
f_{\lambda, \lambda'} (\bsk, \bsq)= \frac{1}{2} (1+\lambda \lambda'\cos(\theta_k-\theta_{k'})).
\label{ff}
\eearr
for $\lambda=-1,\lambda'=+1$. 
Again to avoid cluttering up with notation, we define an auxiliary frequency 
\be
  \Omega\equiv \hbar \omega- \hbar v_x q \eta \cos\phi,
  \label{auxfreq.eqn}
\ee
and we work in units where $\hbar=v_x=1$. 
Furthermore according to the fluctuation dissipation theorem for Eq.\eqref{paiud.eqn},
$\chi_0(\boldsymbol{q},-|\omega|)=\chi^*_0(-\boldsymbol{q},|\omega|)$, which implies, 
$\chi_0(\bsq,|\Omega|)=\chi^*_0(-\bsq,-|\Omega|)$. Hence, we only need to evaluate the integrals
for $\Omega>0$. 
Doing integration on momentum space and using Kramers-Kronig dispersion relation (for more detail see Appendix A) 
gives the following result for imaginary and real part of undoped polarization function.
 \bearr
 &&\Im\chi_0(q,|\Omega|)=-\frac{g q^2 }{16 \hbar v_xv_y } \frac{  {\rm sgn}(\Omega) }{\sqrt{\Omega^2-q^2}} \Theta(|\Omega|-q),
 \label{imun.eqn}
\\
 &&\Re\chi_0(q,|\Omega|)=-\frac{g q^2 }{16 \hbar v_xv_y } \frac{\Theta(q-|\Omega|)}{\sqrt{q^2-\Omega^2}}.
 \label{realun.eqn}
 \eearr
 Here, the functional form of the real and imaginary part are same as the undoped graphene~\cite{Wunsch2006}.
However, the tilt induced direction dependence is encoded in the definition of  $\Omega=\omega-q\eta\cos\phi$. 
Note that only for undoped graphene the entire tilt dependence enters into the auxiliary frequency $\Omega$. 
As will be shown in the following, in the case of doped tilted Dirac cone, the tilt-dependence will not appear
through $\Omega$~\cite{Agarwal2017}, but will also heavily affect the integration limits. 

\begin{figure}[t]
   \includegraphics[width = .47\textwidth]{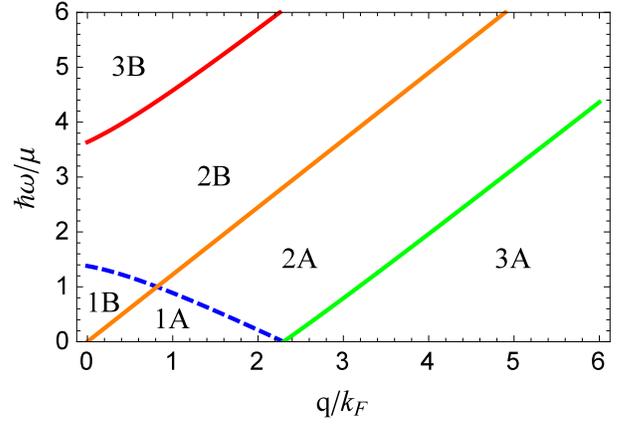} 
   \caption{Different regions in the space of $q/k_F$ and $\hbar\omega/\mu$. Various regions determine 
   sign structure coming from step functions in the tilted Dirac cone. This figure is produced for the tilt parameter
   $\eta=0.45$ which will be extensively used in this paper. In Ref.~\onlinecite{Nishine2010} similar figure
   is produced for $\eta=0.8$. In the limit $\eta\to 0$, this figure becomes identical to Fig. 6 of Ref.~\cite{Wunsch2006}.
   Part of the dashed border that separates region 1B and 2B is where the plasmon kink develops (see sec.~\ref{tilt.sec} of the main text). }
    \label{regions.fig}
\end{figure}

\subsection{Doped tilted Dirac cone}
In the case of doped tilted Dirac cone as a result of nonzero chemical potential ($\mu\neq0$), both process of 
intra- and inter-band transition contributes to the polarization function~\cite{Wunsch2006}. 
As in the case of graphene, it turns out to be more convenient to subtract the polarization of undoped case from doped one, 
\bearr
 \label{subtract}
 &&\Delta \chi(\boldsymbol{q},\omega)= \chi(\boldsymbol{q},\omega)- \chi_0(\boldsymbol{q},\omega)\nn\\
 && \chi(\boldsymbol{q},\omega)=\Delta \chi(\boldsymbol{q},\omega)+\chi_0(\boldsymbol{q},\omega),
\eearr
Then the undoped contribution, $\chi_0$, can be added at the end. 
As pointed out, the subscribe $0$ stands for undoped case. After doing some simplification, $\Delta \chi(\boldsymbol{q},\omega)$ can be rewrite as,
\be
\Delta \chi(\boldsymbol{q},i\Omega_n)=
 \frac{g \gamma^2}{A \hbar^2 v_x^2} \sum_{\bsk,\lambda}   
 \Theta(\mu-\lambda E_{\lambda}(\bsk))P(\bsk,\bsq,i\lambda \Omega_n),
 \label{paid}
 \ee
where,
\bearr
P(\bsk,\bsq,i\Omega_n) =\frac{(i\Omega_n+ k)+ k' m}{(i\Omega_n+ k)^2-(k')^2}.
 \eearr
Here, $k'=|\bsk+\bsq|$ and  $m=\cos(\theta_k-\theta_{k'}) $. 
The function $P(\bsk,\bsq,\pm i\Omega_n)$ is a complex function.
The essential point in the Eq.~\eqref{paid} is that, in comparison to the graphene, the step function not only is a function of $k$, but also
is dependent on the direction of wave vector $\bsk$, which makes the integration more complicated. 
At the end, we need to perform the Wick rotation, $i\Omega_n\to \Omega+i\ep$.  

In what follows, in order to calculate the real part of the polarization function for the doped tilted Dirac cone, 
the infinitesimal imaginary part of $i\ep$ can be ignored, and the integration on $k$ becomes a Cauchy principal value. 
Doing the integration on $k$, generates a logarithmic function the branch cut of which needs to be carefully handled. 
This makes the angular integration slightly complicated. Our trick to overcome this difficulty is to represent the logarithm itself as 
an integral over some auxiliary variable. Then the calculus of residues can be used to perform the angular integration. 
The integral over the auxiliary variable can be calculated at the end. 
For details of calculation see Appendix B. The final result of integration is summarized as,
\bearr
\Re\Delta\chi(\bsq,\Omega)=&& F^1(\bsq,\Omega) [G(X^+)|_{x_d^+}^{x_u^+} +\sum G(X^-)|_{x_d^-}^{x_u^-}]\nn\\&&+F^0(\bsq,\Omega),
\label{realdoped}
\eearr
where,
\bearr
F^0(\bsq,\Omega)&=&\frac{g}{16\pi \hbar^2 v_x v_y } \frac{\mu ~q^2}{ \sqrt{1-\eta^2}} \frac{ A(\bsq,\Omega)}{ D^2(\bsq,\Omega)},
\\
F^1(\bsq,\Omega)&=&\frac{g}{16\pi \hbar^2 v_x v_y } \frac{q^2}{  \sqrt{|\Omega^2-q^2|}},\\
G(x)&=&B(\bsq,\Omega) x \sqrt{x^2-x'}\nn\\&&
-{\rm sgn}(\Omega-q)  \cosh^{-1} \frac{x}{\sqrt{x'}},
\eearr
and the summation denoted by $\Sigma$ indicates sum over disconnected pieces.
The quantities $x'$ and $X^\pm$ are defined as follows,
$x'=(\Omega+q \eta \cos\phi)^2-(1-\eta^2)(\Omega^2-q^2)$, and 
$X^{\lambda}=2\tilde{\mu} x+ \lambda (\Omega+q\eta \cos\phi)$. The $\tilde\mu$ and
upper ($x_u^{\lambda}$) and lower ($x_d^{\lambda}$) limits for $\lambda=+(-)$ corresponding to intra- (inter-) band processes,
are limits of integrations which are separately determined for each region in Fig.~\ref{regions.fig}. 
The details of the derivations of various regions are given in Appendix B. 
Here ${\rm sgn}(\Omega-q)$ is the sign of $\Omega-q$. The definitions of coefficients $A(\bsq,\Omega)$, $B(\bsq,\Omega)$, and $D(\bsq,\Omega)$
are given in Appendix B. These function strongly depend on the tilt parameter $\eta$, and hence on the direction $\phi$ of momentum transfer $\bsq$. 
Therefore the tilt and angular dependence in doped Dirac cone not only comes through the auxiliary frequency, $\Omega$, 
but it also appears in coefficient of $A , B , D$. The later part is missed in the calculation of Ref.~\onlinecite{Agrawal}. 

As a cross-check of our analytic results against the established results on graphene~\cite{Wunsch2006,hwang2007}. 
If we set the tilt parameter $\eta=0$, it can be easily seen that the above functions reduce to, $A(\bsq,\Omega)=q^2$, $B(\bsq,\Omega)=q^{-2}$. As a result, the real part Eq.~\eqref{realdoped} reduces to the real part of doped graphene polarization function~\cite{Wunsch2006}.

Now let us look into the simpler calculation which deals with the imaginary part of polarization function Eq.~\eqref{paid}~\cite{Nishine2010}. 
This can be straightforwardly calculated thanks to a Dirac delta function arising from small imaginary part $i\ep$ in the
denominator of Eq.~\eqref{paid}. The imaginary part of the polarization function in our notation becomes, 
\bearr
\Im\Delta\chi(\bsq,\Omega)= F_2(\bsq,\Omega) [G^{+}_0(x)|_{y_d^+}^{y_u^+} +G^{-}_0(x)|_{y_d^-}^{y_u^-}],
\label{imdoped}
\eearr
where, 
\bearr
&&F_2(\bsq,\Omega)=\frac{g}{32\pi \hbar^2 v_x v_y } \frac{q^2}{ \sqrt{|\Omega^2-q^2|}},
\nn\\
 &&G_0^+(x)= x \sqrt{x^2-1}-  \cosh^{-1}x,\nn\\&&
 G_0^-(x)= x \sqrt{1-x^2}+ \sin^{-1}x.
\eearr
Here the upper ($y^\lambda_u$) and lower ($y^\lambda_d$) limits are defined by the roots of Fermi distribution (step function at zero temperature). 
Their explicit expressions are given in Appendix C.  

Again it can be seen (see Appendix C) that in the case of $\eta=0$, the imaginary part as well, 
will be reduced to the case of graphene.
Finally, as the last step, we should add the undoped polarization function  to the $\Delta\chi(\bsq,\Omega)$ to derive the doped polarization function. 

\begin{figure}[t]
   \includegraphics [width = .47\textwidth]{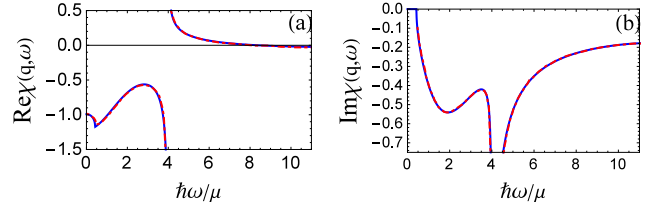}
   \caption{Comparison of real and imaginary part of polarization function for doped tilted Dirac cone with the result in Ref.~\onlinecite{Nishine2010}. 
   The vertical axis is in units of $\mu/\hbar^2 v_x v_y$ and $v_{x0}=0.8$ and $v_{y0}=0$ i.e. $\eta=0.8$. The momentum $q=4$ and is in the $y$ 
   direction ($\phi=\pi/2$). The blue (solid) and red (dot dashed) lines curves correspond to our calculation and Ref.~\onlinecite{Nishine2010} result, respectively.}
    \label{suzumura.fig}
\end{figure}

\subsection{Benchmark against existing results}
Nishine and coworkers have already obtained the analytical formula for the imaginary part~\cite{Nishine2010}. For the
real part, they numerically perform the Kramers-Kronig transformation. Therefore they have numerical results 
for the Re$\chi(\bsq,\omega)$. Let us ensure that our analytic results agree well with their results. 
In the following, we reproduce some of the plots related to their work. 
In Fig.\ref{suzumura.fig}-(a) we compare our analytic Eq.~\eqref{realdoped}, for the real part of polarization (solid blue curve) 
with the numerical result of Nishine and coworkers (red dot dashed). 
As can be seen, the agreement is perfect. 
Panel (b) of this figure compares the
imaginary parts adopted from their curves with those produced by our Eq.~\eqref{imdoped}. The comparisons are made for
$\eta=0.8$ used in Ref.~\onlinecite{Nishine2010}. Also in both cases the vertical axis is in units of 
$\mu/\hbar^2 v_x v_y$. Again, as can be seen, the agreement is perfect. Our analytical formula for the real part will 
allow us to analytically explore the plasmons and screening in tilted Dirac cone. In the following section, we 
start with a qualitative discussion of plasmons in presence of tilt.  

\section{Plasmons: Role of tilt parameter $\eta$}
One of the significant collective excitations of the electronic systems in long wavelength limit is plasmon,  which argument
the single particle picture of an electron gas at lattice scale with a self-organized collective oscillations of 
appropriate electric fields and charge densities~\cite{Phillips,BohmPines1953}. 
In two dimensional electron gas, whether Dirac~\cite{castro} or non-Dirac, the plasmon 
dispersion relation is at the long wavelength limit is given by $\omega_{\rm pl}\propto\sqrt{q}$.  Indeed this follows
from a general hydrodynamic consideration~\cite{Fetter1973}. In the linear response formulation, plasmons are zeros of the dielectric function. 
Within the RPA approximation, the dielectric function will be given by, 
\be
   \varepsilon(\bsq,\omega)=1-V_q \chi(\bsq,\om),
   \label{dielectric.eqn}
\ee
where, in a single layer of the two-dimensional system, the Coulomb interaction is given by $V_q=2\pi e^2/q$, and $\chi$ is the 
bare electron-hole bubble. Since $V_q$ is a positive quantity, a necessary condition to obtain a plasmon branch of excitations is
that the real part of density response function be positive. 

\begin{figure}[t]
   \includegraphics[width = .47\textwidth] {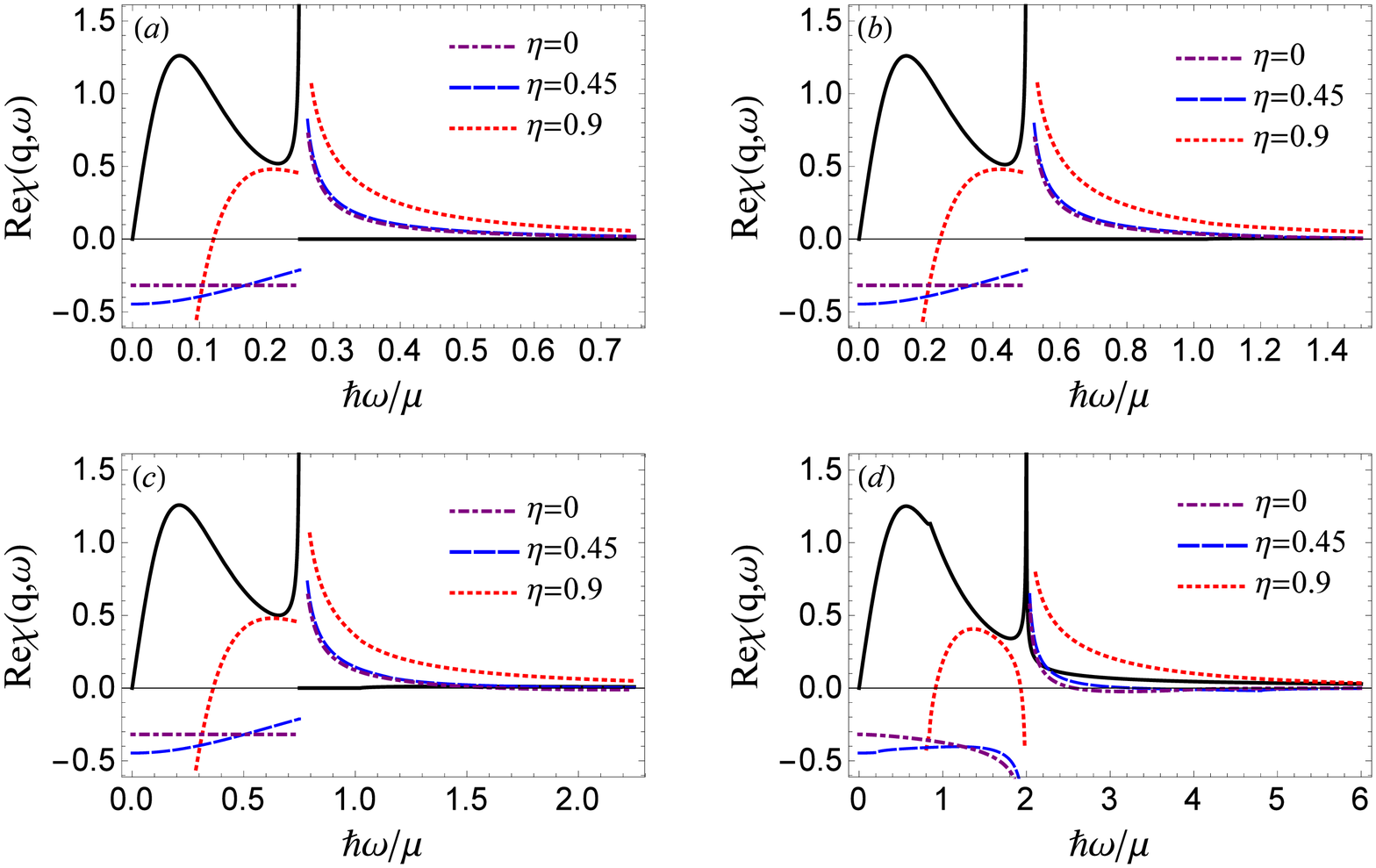}
   \caption{Real part of polarization function for doped tilted Dirac cone, for different value of $\eta$ as indicated
   in the legend. The direction of $\bsq$ is fixed by $\phi=\pi/2$. The vertical axis is in the unit  $\mu/\hbar^2 v_x v_y$ and the 
   horizontal axis is a dimensionless quantity $\hbar \omega/\mu$. The magnitude of $q/k_F$ in various panels are, $0.25$ in (a), $0.5$ in (b),
   $0.75$ in (c), and $2.0$ in (d). Solid line is the imaginary part of $\chi$ for $\eta=0.9$. For explanation see the text. }
    \label{difeta.fig}
\end{figure}

In Fig.~\ref{difeta.fig} we have plotted real part of polarization function, $\chi$ (in the unit  $\mu/\hbar^2 v_x v_y$) for different value of $\eta$.
The dot dashed, dashed, and dotted plots correspond to $\eta=0,0.45,0.9$, as indicated in the legend. Four panels (a), (b), (c), and (d), 
correspond to $q=|\bsq|=0.25,0.5,0.75$, and $2.0$, respectively. 
Since the polarization function is strongly anisotropic, in this figure we have fixed the direction $\phi$ of $\bsq$ to be at $\phi=\pi/2$.
The solid line is the plot of imaginary part of $\chi$ for $\eta=0.9$ only. The above collective mode equation in units of $\hbar=v_x=1$
can be written in the dimensionless form, Re$\chi=\bar q v_y/(\alpha c)$, where $\alpha=1/137$ is the fine structure constant, $c$ is the
velocity of light, and dimensionless in their natural units. For example, $q$ is meant in units of $k_F$.
Quite generally, the imaginary part of $\chi$ abruptly changes at $\omega_+=q(1+\eta\cos\phi)$, which marks the upper
border of intra-band PH excitations in the tilted Dirac cone~\cite{Nishine2010}. For $\phi=\pi/2$ this reduces to $\omega_+=q$.
This is why in both the real part and the imaginary part (solid line) there is a discontinuity at this energy scale
which for $\phi=\pi/2$, coincides with $q$ itself. 
The plasmon mode is obtained by intersecting a constant horizontal line (given by the above dimensionless equation)
with the real part of $\chi$. Let us first focus on $\omega>\omega_+$ region (1B in Fig.~\ref{regions.fig}), 
where the Im$\chi$ is identically zero.  As can be seen in all figures, by increasing 
the tilt parameter $\eta$, the real part of $\chi$ is lifted to larger values. This ,in turn, will shift the
plasmon modes to higher energies. Therefore the generic effect of the increase in the tilt is to shift the
plasmons to higher energies. The $\eta=0$ would correspond to the graphene-like situation. This is the standard
plasmon branch. This branch will continue to 2B region of Fig.~\ref{regions.fig}, but will acquire small damping as there are 
small density of inter-band PH excitations in 2B. 

\begin{figure}[t]
   \includegraphics[width =0.49\textwidth]{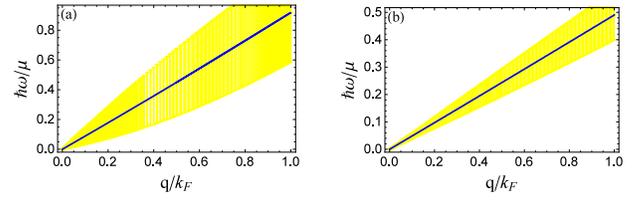}
   \caption{(Color online) The overdamped plasmon mode arising from tilt for the direction $\phi=\pi/2$. 
   Left (right) panel corresponds to tilt parameter $\eta=0.75$ ($\eta=0.9$). 
   The shaded region represents damping. 
   The amount of damping in the right panel is decreased by a factor of $10$ to fit in the panel. 
   }
   \label{overdamped.fig}
\end{figure}

Now let us look into $\omega<\omega_+$ (corresponding to region 1A in Fig.~\ref{regions.fig}), where the imaginary part is non-zero and
non-negligible. 
It is curious to note that for large values of $\eta$, (the red dotted line in all panels), the real part of $\chi$
for $\omega<\omega_+$ becomes positive. This implies a lower energy plasmon branch. However, since in the natural
units, the magnitude of the imaginary part -- which quantifies the density of free intra-band PH states 
(black, solid line) available for
Landau damping -- is ${\cal O}(1)$, such a tilt induced extra plasmon branch will be over-damped. 
The dispersion of overdamped plasmon branch is shown in Fig.~\ref{overdamped.fig}. 
The shaded region indicates the damping. larger width means larger damping~\cite{SaharStoner}. As can be seen
the mode disperses linearly, but it is heavily damped. In the right panel corresponding to $\eta=0.9$ the
damping is so large that, in order to fit in the panel, we have reduced the shaded region indicating the damping by a factor of $10$. 
Note that for 
$\eta=0$ situation pertinent to graphene, the real part for $\omega<\omega_+$ can never be positive, and hence
no extra plasmon solution is conceivable. 

\begin{figure}[b]
   \includegraphics[width = .47\textwidth] {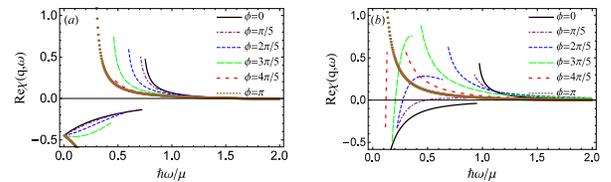}
   \caption{Real part of $\chi$ in units of $\mu/\hbar^2v_xv_y$ as a function of dimensionless $\omega/\mu$ for 
   fixed $q/k_F=0.5$ and various values of angle $\phi$ indicated in the legend. The left (right) panel 
   corresponds to $\eta=0.45$ ($\eta=0.9$).} 
   \label{angular-dep.fig}
\end{figure}

The above discussions in Fig.~\ref{difeta.fig} was for a fixed $\phi=\pi/2$ orientation of $\bsq$.
Let us now explore the direction dependence. 
Since the density response  is strongly anisotropic, in Fig.~\ref{angular-dep.fig} we have shown the angular dependence of 
Re$\chi$. Solid (black), dot dashed (purple), dashed (blue), long dashed (green), sparse dashed (red), and dotted (brown)
correspond to $\phi=n\pi/5$ for $n=0,1,2,3,4,5$, respectively. The left (right) panel corresponds to the tilt parameter $\eta=0.45$ ($\eta=0.9$).
The first observation concerns the $\omega>\omega_+$ (1B in Fig.~\ref{regions.fig}) region. As can be seen by increasing $\phi$ from $0$ to $\pi$ in both panels, 
the Re$\chi$ curves are pushed to the left as $\omega_+$ itself depends on the angle $\phi$. Therefore the corresponding
plasmons will have smaller energies. The second observation which is the essential difference between the left ($\eta=0.45$) and
right ($\eta=0.9$) panel is concerned with the $\omega<\omega_+$ (1A in Fig.~\ref{regions.fig}) region. 
As can be seen for $\eta=0.9$ in the right panel, for all angels, except $\phi=0$, the Re$\chi$ develops a positive branch
which gives rise to over damped plasmon in $\omega<\omega_+$ (1A in Fig.~\ref{regions.fig}) region. 
This indicates that the additional over damped plasmon branch is solely due to (large enough) tilt of the Dirac cone. 

\begin{figure}[t]
   \includegraphics[width =0.45\textwidth]{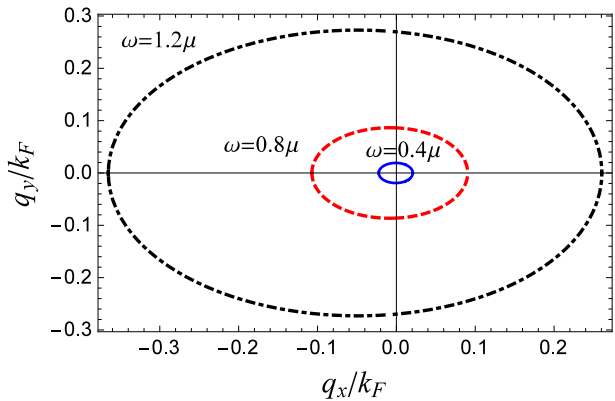}
   \caption{(Color online) Constant energy cuts of the plasmon dispersion for a moderate 
   tilt parameter $\eta=0.45$. Curves are generated for $v_x=v_y=c/1000$.
   }
   \label{polar.fig}
\end{figure}

Quite generally the anisotropy can come from two sources: (i) the intrinsic anisotropy due to
$v_x\ne v_y$, or equivalently $\gamma\ne 1$. (ii) the tilt also acts as a source of anisotropy which 
is manifested in Fig.~\ref{angular-dep.fig} as a strong angular dependence of the Re$\chi$-vs-$\omega$ curves. 
To investigate this further, in Fig.~\ref{polar.fig} we have plotted the constant energy contours of the plasmon dispersion
for a fixed tilt parameter $\eta=0.45$, for three representative energies as indicated in the figure. 
The contours clearly indicate a tilt pattern. Moreover, it is manifestly symmetric with respect to $\phi\to -\phi$,
which is expected from the Hamiltonian, as we have assumed the tilt is along $k_x$ axis.  
The plasmonic energy contours in Fig.~\ref{polar.fig} reflect the sole effect of tilt 
parameter, as we have generated this figure for $v_x=v_y$. 
When the tilt parameter is set to zero, the above ellipses become concentric,
and the aspect ratio becomes, $1$, meaning that the ellipses become circles.

\section{Tilt-induced kink in the plasmon dispersion}
\label{tilt.sec}
In the case of graphene where the tilt parameter, $\eta$ is zero,  the region 1B of Fig.~\ref{regions.fig} reduces to a triangular
region which is void of free PH pairs, and separates the intra-band (lower side)
continuum of PH excitations from the inter-band (upper side) continuum. The plasmons
in region 1B are well defined. The plasmon branch however, continues inside the region 2B of 
Fig.~\ref{regions.fig} which contains very small amount of DOS of inter-band 
PH excitations. Therefore the plasmon branch continuously enters the inter-band PHC with a 
negligible damping~\cite{hwang2007}. By turning on the tilt parameter, $\eta$, the density of 
inter-band PH states in region 1B will not appreciably change. Therefore the plasmon branch
will continue to the region 1B with negligible damping. But as we will see in this section, 
the tilt parameter will induce
a kink at the border separating regions 1B and 2B (dashed line in Fig.~\ref{regions.fig}).

\begin{figure}[t]
   \includegraphics[width =0.50\textwidth]{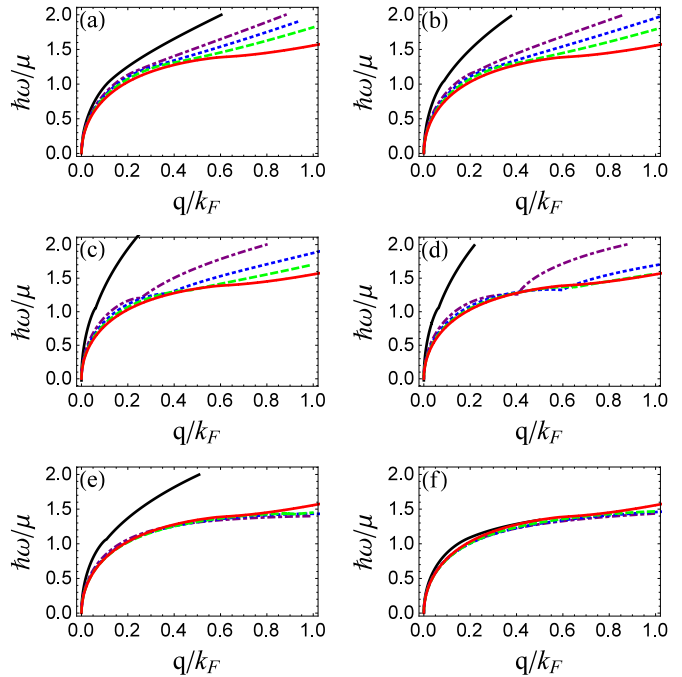}
   \caption{(Color online) Dependence of plasmon dispersion on tilt parameter $\eta$ and angle $\phi$. 
   The horizontal axis is $q/k_F$ and vertical axis is $\hbar\omega/\mu$. The tilt parameters
   $\eta=0,0.3,0.45,0.6,0.9$ are encoded as, solid (red), dashed (green), dotted (blue) dot-dashed (purple) 
   and the leftmost solid (black) curves, respectively.  Panels (a)-(f) correspond to
   directions $\phi=\pi n/5$, with $n=0,\ldots,5$.}
   \label{kink.fig}
\end{figure}

Let us start by monitoring the evolution of plasmon branch as the tilt parameter grows from zero. 
In Fig.~\ref{kink.fig} we have plotted the dispersion of plasmons in the plane of $\hbar\omega/\mu$ and $q/k_F$. 
Panels (a)-(f) corresponds to angles $\phi=n\pi/5$ with $n=0,\ldots,5$. Various curves as indicated in 
the legend correspond to tilt parameters $\eta=0,0.3,0.45,0.6,0.9$. 
The first point to notice is the following:
A common aspect of all panels (all directions) in Fig.~\ref{kink.fig}
is that in the long wavelength limit for a fixed small $q$, the energy of the plasmon resonance
increases by increasing the tilt parameter $\eta$. This is true for all angles in panels (a) to (f). 
Such ordering in the energy of plasmon resonances in terms of $\eta$ does not hold 
for larger $q$ values, anymore. The second point
to notice is that for $\phi=\pi$ in panel (f), the plasmon dispersion is less sensitive to the tilt parameter $\eta$. 

\begin{figure}[t]
   \includegraphics[width =0.41\textwidth]{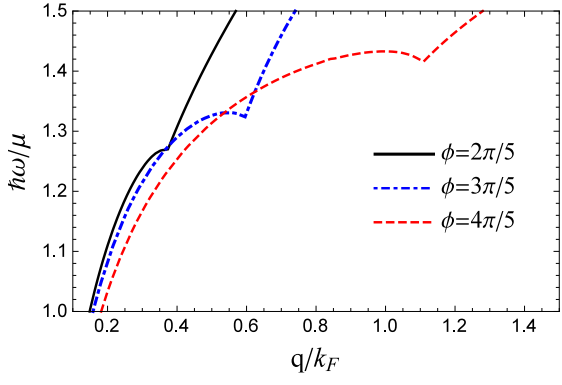}
   \caption{(Color online) Angular dependence of the plasmonic kink. The figure is produced
   for the moderate tilt parameter $\eta=0.45$, pertinent to realistic materials~\cite{Goerbig2014}. 
   }
   \label{kink-eta-fixed.fig}
\end{figure}

The third and most remarkable point to notice is the appearance of a kink in the plasmon dispersion.
This is more manifest in panels (c) and (d) corresponding to $\phi=2\pi/5$ and $\phi=3\pi/5$, respectively. 
Indeed the most manifest form of kink appears for $\phi=\pi/2$. 
The kink is present for any non-zero 
tilt parameter $\eta$. Therefore {\em the above anisotropic kink is a direct manifestation of
the tilt.} This fact can be used to directly map the tilt parameter from the angle-resolved electron energy loss 
spectroscopy (EELS). The kink is very anisotropic. To bring this out, 
in Fig.~\ref{kink-eta-fixed.fig}, for a fixed moderate value of $\eta=0.45$~\cite{Goerbig2014}, 
we have plotted the plasmon dispersion for various angles. As can be seen, the position of kink is very sensitive to
the direction $\phi$ of the momentum $\bsq$ of the plasmon excitations. 

\subsection{Origin of the kink and direct experimental measurement of the tilt parameter}
\begin{figure}[t]
   \includegraphics[width =0.47\textwidth]{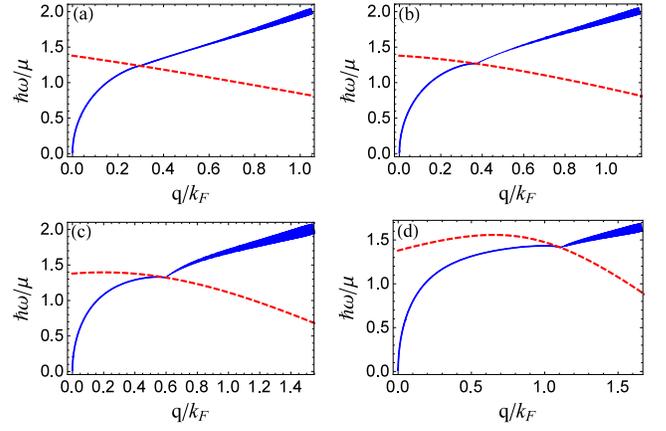}
   \caption{(Color online) The position of the kink for the tilt parameter $\eta=0.45$.  
   The Solid (blue) line is the plasmon dispersion, and the dashed (red) curve is the 
   $\omega_A(q)$ given by Eq.~\eqref{omega_A.eqn}. Panels (a)-(d) correspond to $\phi=n\pi/5$
   with $n=1,2,3,4$, respectively. Note that {\em the width of the plasmon dispersion in region 2B 
   is exaggerated by a factor of $50$ to emphasize very small damping.}
   }
   \label{position.fig}
\end{figure}

To investigate the origin of kink, in Fig.~\ref{position.fig} we have plotted
the plasmon dispersion (the solid, blue, curve) for the tilt parameter $\eta=0.45$. Panels (a) to (d) 
correspond to directions $\phi=n\pi/5$ with $n=1,2,3,4$, respectively. 
The thickness is associated with the damping of plasmons. To be clear, we have exaggerated the
thickness by a factor of $50$. This clearly indicates that the emergence of kink goes hand in hand with the
onset of damping. Therefore the kink appears at the border separating regions 1B and 2B of Fig.~\ref{regions.fig}. 
To verify this, we have plotted the border formula by dashed (red) line. As is expected the kink begins exactly when the
plasmon branch crosses this border. 

The formula for the dashed border which now with hindsight can be dubbed {\em kink energy scale} 
is given by~\cite{Nishine2010},
\be
   \omega_{\rm kink}=q\eta\cos\phi+\frac{2}{1-\eta^2}-\sqrt{q^2+\frac{4 q\eta \cos\phi }{1-\eta^2}+\left(\frac{2\eta}{1-\eta^2}\right)^2}.
   \label{omega_A.eqn}
\ee
The above energy scale is denoted by $\omega_A$ in Ref.~\onlinecite{Nishine2010}, 
and defines the upper border of the region denoted as 1B in Fig.~\ref{regions.fig}. 
This region is defined in 
\be
    q(1+\eta\cos\phi)<\omega<\omega_{\rm kink},~~~q<1
\ee
It can be easily checked that, for $\eta=0$ this region reduces to the triangular region 
that separates the inter-band and intra-band portion of PHC in doped graphene~\cite{Wunsch2006,hwang2007}.
When the borders coincide with that of the triangular 1B region of the $\eta=0$ situation, the parts of
the plasmon branch in 1B and 2B region connect to each other without any kink~\cite{hwang2007}.
However, the tilt pushes down the upper boundary of the triangle and distorts it to the dashed 
curve in Fig.~\ref{regions.fig}, whereby a kink in the dispersion of plasmon branch is generated. 

The above kink in plasmon dispersion allows for a direct measurement of the tilt parameter. 
As pointed out, beyond the kink that the plasmon dispersion enters the continuum of inter-band PH excitations, the
imaginary part of the polarization function $\chi$ is negligibly small (the width of the plasmon in Fig.~\ref{position.fig}
is exaggerated by a factor of $50$ to emphasize the connection between damping and kink)
Therefore the plasmon excitations with wave-vectors larger than the kink scale
are expected to live long enough to allow for experimental detection. 
Suppose that for a fixed direction $\phi_{\rm exp}$, the kink is experimentally determined to occur at $(q_{\rm exp},\omega_{\rm exp})$.
Then the tilt parameter $\eta$ satisfies the following equation,
\be
   \omega_{\rm exp}=\omega_{\rm kink}(q_{\rm exp},\phi_{\rm exp},\eta)
   \label{emperical_eta.eqn}
\ee
where the $\omega_{\rm kink}$ function in the right hand side is given by Eq.~\eqref{omega_A.eqn}. 
All quantities with the suffix "exp" can be directly measured in the experiment. 
Therefore the only unknown in the Eq.~\eqref{emperical_eta.eqn} is the tilt parameter $\eta$. 
Therefore, this equation enables a {\em direct} experimental measurement of the tilt parameter, $\eta$.

\section{Asymptotic formula}
The hydrodynamic limit in electron liquids is quite generally given by $\omega\to 0$ and $q\to 0$.
Their ratio however, can be finite. We are working in units where the velocity $v_x=v_y=1$, and $\hbar=1$.
Hence the ratio $\omega/q$ is dimensionless. The tilt parameter, $\eta$, being dimensionless, enters 
the game through a combination $\omega\eta/q$. Let us see this by asymptotically expanding our
analytic formula for the Re$\chi$. 

\subsection{Tilt-dependence of plasmons}
In order to investigate the plasmon dispersion in tilted Dirac cone, we first find the long wavelength limit of polarization function.
As pointed out, the tilt parameter appears as the $\omega\eta/q$ combination. 
In the long wavelength limit ($q\rightarrow0$) our formula gives,
\bearr
&& {\rm Re}\chi(q\rightarrow0,\omega)\approx\nn\\&& \begin{cases}
    \frac{D_0 q^2}{4\pi \omega^2} (1-\frac{2\omega\eta}{q}),&  \eta\ll q ~~ ,~~ \frac{\omega \eta }{q}\ll 1  \\
     \frac{D_0q^2 }{4 \pi\omega^2 \eta^2}\left[\cos2\phi+H(\eta)\right],& \eta\gg q ~~,~~ \frac{\omega \eta }{q}\gg 1  
 \label{longwl.eqn}
     \end{cases}
\eearr
where,
\be
H(\eta)=\frac{\eta^2+(\eta^2-2)\cos2\phi}{\sqrt{1-\eta^2}}~~~,~~~D_0=\frac{g \mu}{4\pi \hbar^2 v_x v_y},
\ee
where we have restored the constants $\hbar,v_x,v_y$. 
The $\eta\ll q$ piece is continuously connected to $\eta\to 0$. Indeed setting $\eta=0$ in the first piece of the above
function, we obtain the correct expression for graphene~\cite{Wunsch2006}. Therefor one recovers the standard
plasmon dispersion of graphene given by~\cite{Wunsch2006,hwang2007}
\be
   \hbar\omega_{\rm pl}=\hbar\omega_0 \sqrt{q},
   \label{plgraphene.eqn}
\ee
where $\omega_0=\sqrt{g\mu e^2/(8\pi v_x v_y)}$ is set by doping, $\mu$, and Coulomb interaction $e^2$. 
As pointed out, the $\sqrt{q}$ plasmonic dispersion is a generic characteristic of 2D systems and can be obtained from
hydrodynamic treatment~\cite{Fetter1973}. 

Now let us look at the other limit where $q$ is still very small, but $\eta$ is finite such that the combination $\omega\eta/ q$ is very large. 
In this regime, the second piece of Eq.~\eqref{longwl.eqn} determines the behavior of plasmons. 
Therefore, the plasmon dispersion is given by,
\be
   \hbar\omega_{\rm pl}=\hbar\omega_\eta \sqrt{q},~~~~~~
   \omega_\eta=\omega_0\frac{\sqrt{\cos2\phi+H(\eta)}}{\eta}.
   \label{pltilt.eqn}
\ee
Note that due to the piecewise nature of Eq.~\eqref{longwl.eqn}, the $\eta= 0$ in Eq.~\eqref{pltilt.eqn}
does not reduce it to Eq.~\eqref{plgraphene.eqn}. The $\eta=0$ limit is consistent with the first piece of 
Eq.~\eqref{longwl.eqn}. 
As can be seen in the case of tilted Dirac cone dispersion, in addition to dependence on the tilt parameter, $\eta$, 
there is a substantial dependence on the angle $\phi$ of the wave vector $\bsq$. 
It is not surprising that the presence of tilt does not change the generic $\sqrt q$ dependence of the plasmon
oscillations in a single layer, as it follows from quite general hydrodynamic arguments~\cite{Fetter1973}.

\subsection{Tilt enhances the static screening}
Now let us look at the opposite limit where $\omega$ is set to zero. The static polarization
contains information about the screening of external charges in the ground state. 
In the limit of low frequency setting $\omega=0$ implies $\Omega=-q \eta \cos\phi$. 
Notice that depending on the orientation $\phi$ of the wave vector $\bsq$ with respect
to $x$ axis, the auxiliary frequency $\Omega$ can be either positive or negative quantity. 
From the general expression in Eq.~\eqref{realdoped} for negative $\Omega$ (corresponding to $\cos\phi>0$) 
we can find the following representation of polarization  for any $q$ and $0 < \phi < \pi$,
\bearr
{\rm Re}\chi(0,\bsq)=-\frac{g \mu}{2 \pi \hbar v_x v_y \sqrt{(1-\eta^2)^3}}[1+\Theta(\tilde{q}-2\mu) f(\frac{\tilde{q}}{2\mu})],\nn\\
\label{screening}
\eearr
where we have restored the constants $\hbar, v_x, v_y$.
The $\tilde q$ is defined as, $\tilde{q}=q\sqrt{(1-\eta^2)(1-\eta^2\cos^2\phi)}$, and 
\bearr
f(\frac{\tilde{q}}{2\mu})=&&-\frac{q^2}{2\tilde{q}^2}\sqrt{1-(\frac{2 \mu}{\tilde{q}})^2}(1-\eta^2)(1-\eta^2\cos2\phi)\nn\\&&+\frac{q^2}{4\tilde{q} \mu}\cos^{-1}(\frac{2\mu}{\tilde{q}}).
\eearr
Again the first point to note is that for $\eta\to 0$, the above expression reduces to the corresponding
expression of graphene~\cite{Wunsch2006,hwang2007,Katsnelsonbook}. The second point to notice is that, 
the screening is controlled by doping $\mu$, and will be ineffective when $\mu=0$, as there will be
no single particle DOS at the Fermi level to take care of screening. This argument holds for any
tilt parameter, in agreement with Eq.~\eqref{screening}. The third point to notice is the strong direction 
dependence of screening. 

To manifestly see the role of tilt in screening, let us consider a very special regime corresponding to
$\tilde{q}<2\mu$. In this situation, the second term in Eq.~\eqref{screening} goes away, and  
above relation will become $q$-independent. Then it can be easily seen that the screening is
controlled by $\mu (1-\eta^2)^{-3/2}$. In the $\eta\to 0$ limit it reduces to the standard formula
of graphene, while for $\eta\to 1$, the above quantity diverges. Therefore as far as static screening
is concerned, the role of very large $\eta\lesssim 1$ is to effectively enhance $\mu$ according to the
above formula. Therefore for a fixed doping level, the larger tilt $\eta$ is expected to give rise to 
better screening. 

\section{Summary and conclusion}
In this paper, we obtained analytic representation of the full polarization function
for tilted two-dimensional Dirac cone with arbitrary tilt parameter, $0\ge \eta<1$ and
for arbitrary doping. Our formula agrees with the numerical evaluation of the real part of 
polarization in Ref.~\onlinecite{Nishine2010} that numerically implements Kramers-Kronig relation. 
Our result does not agree with Ref.~\onlinecite{Agrawal}, and the reason is that the tilt parameter, 
affects the results both through the auxiliary frequency, Eq.~\eqref{auxfreq.eqn}, {\em and through the limits
of integration}. 

Our analytic formula enables us to find -- in addition to a trivial tilt-induced anisotropy -- two interesting
plasmonic effects. First of all, for strong enough tilt, a new branch of over-damped plasmon appears 
which overlaps with a large DOS of intra-band PH excitations. When the tilt parameter is zero or even
small, there is no chance for such an overdamped plasmonic mode that disperses linearly. Since this overdamped mode energetically
overlaps with intra-band PHC, it is expected to affect the single-particle properties. 
Secondly, the upper boundary of the region 1B in Fig.~\ref{regions.fig} is strongly modified
by the tilt. This modification gives rise to a kink in the plasmonic dispersion, the position 
of which is right at the dashed border in Fig.~\ref{regions.fig} that separate 1B and 2B regions.
In region 2B the mode acquires a negligible damping which is due to its overlap with inter-band
PH excitations. The small damping allows the branch in region 2B to live long enough to allow for
experimental detection of the kink in the plasmonic dispersion. 

The analytic formula for the kink energy scale in Eq.~\eqref{emperical_eta.eqn} allows for
a direct experimental measurement of the kink parameter by measuring the wave vector $\bsq$ and
energy $\omega$ at which the kink is taking place. 


\appendix
\begin{widetext}
In these appendices, we provide detailed derivation of the analytic form of the
polarization function for tilted two-dimensional Dirac cone.
\section{Undoped case}
The undoped polarization function Eq.~\eqref{paiud.eqn} consists of two terms.  
The second term is same to the first term if we replace $\Omega\rightarrow -\Omega$. The form factor is defied as
\bearr
F_{\lambda, \lambda'} (\bsk, \bsq)=F_{\lambda, \lambda'} (\bsk, \bsk')=\bra{\bsk,\lambda}\ket{\bsk',\lambda'}= \frac{1}{2} (1+\lambda \lambda'\cos(\theta_k-\theta_{k'})).
\eearr
For simplicity of calculation we change the integration variable  $\theta-\phi\rightarrow \theta$ 
(note that for the polar angle $\theta_k$ of the integration variable $\bsk$ we simply use $\theta$, and
the polar angle of $\bsq$ is $\phi$) and find,
 \be
 \cos(\theta_k-\theta_{k'})= \frac{k +q \cos\theta}{|\bsk+\bsq|},~~~~~~|\bsk+\bsq|=\sqrt{k^2+q^2+2kq\cos\theta}.
 \ee
 
Let us start with the imaginary part which is easier,
 \bearr
 &&\Im\chi_u(\boldsymbol{q},\Omega)=-\frac{g \pi \gamma^2 }{A \hbar v_x^2} \lim\limits_{\ep\rightarrow 0}  \sum_{k} F^{ij}_{-,+}\left(\boldsymbol{k},\boldsymbol{q}\right)   \bigg\{\delta( \Omega-k-|\bsk+\bsq|)  - \delta( \Omega+k+|\bsk+\bsq|) \bigg\}
    .\label{iud.eqn}
 \eearr
Here we replaced $v_x k\equiv k$,  $v_x q\equiv q$, $v_{x0}=\eta v_x \cos\theta_t$, $v_{y0}=\eta v_y \sin\theta_t$ and $\Omega\equiv\omega- q \eta \cos\phi$. 
Furthermore in Eq.~\eqref{iud.eqn} the first (second) term is nonzero whenever $\Omega>0$ ($\Omega<0$). 
By the fluctuation dissipation theory we find $\Im\chi_u(q,|\Omega|)=-\Im\chi_u(q,-|\Omega|)$ or $\Im\chi_u(\boldsymbol{q},|\omega|)=-\Im\chi_u(-\boldsymbol{q},-|\omega|)$. 
So we need to do the integration only for one sign of $\Omega$. 
In the following, we perform the momentum space integration for only the first term  in Eq.~\eqref{iud.eqn}, which gives,
 \bearr
 &&\Im\chi_u(q,|\Omega|)=-\frac{g q^2 }{16 \hbar v_xv_y } \frac{  \sgn(\Omega) }{\sqrt{\Omega^2-q^2}} \theta(|\Omega|-q).
 \eearr
Here we used,
\be
\delta(\Omega-k-|\bsk+\bsq|)=\frac{|\bsk+\bsq|}{k q} \frac{\delta(\theta-\theta_0)}{ |\sin\theta_0|}~~~~~,~~~~~~\cos\theta_0=\frac{\Omega^2-2\Omega k-q^2}{2 k q},
\ee
which given the fact that $-1\leq \cos\theta_0\leq1$, implies $\Omega>q$ and $\Omega-q\leq 2k \leq \Omega+q$.  
Using Kramers-Kronig dispersion relation, the real part will be given by principle integration on domain $\omega'$ as,
  \bearr
  \Re\chi_u(q,\omega)=\frac{p}{\pi} \int\limits_{-\infty}^{+\infty} \frac{\Im\chi_u(\bsq,\omega')}{\omega'-\omega} d\omega'=\frac{p}{\pi} \int\limits_{-\infty}^{+\infty} \frac{\Im\chi_u(q,\Omega')}{\Omega'-\Omega} d\Omega',
   \eearr
By substitution of $\Omega^{'2}-q^2=t^2$, and doing integration on $0\leq t\leq \ep_D$ ($\ep_D$ is energy cutoff which goes away by 
dimensional regularization~\cite{Peskin}) we find, 
 \bearr
 &&\Re\chi_u(q,|\Omega|)=-\frac{g q^2 }{16 \hbar v_xv_y } \frac{\theta(q-|\Omega|)}{\sqrt{q^2-\Omega^2}}.
 \eearr

The above expressions are identical to that of up-right (non-tilted) Dirac cone, except that $\omega \to \Omega=\omega-q\eta\cos\phi$.
This is only true for the undoped tilted Dirac cone. In the doped cases, as we will see, the tilt will heavily affect the limits
of integration. Ignoring this point gives rise to incorrect result~\cite{Agrawal}.

\section{Doped case}
Polarization function of tilted Dirac cone in doped case has a more complicated structure due to the combination of two effects, (i) the angular dependence of 
the single-particle dispersion, (ii) and non zero chemical potential. In order to find the real and imaginary part of polarization we subtract the 
polarization function of undoped system, and then finally after doing the integration on $\bsk$ space, we add it back. 
If we subtract Eq.~\eqref{paiud.eqn} and toggle the integration variables as, 
$\bsk\leftrightarrow -\bsk'$ and $\theta_k\leftrightarrow\theta_{k'}+\pi$ ($\bsk'=\bsk+\bsq$) we find
 \bearr
\Delta \chi(\boldsymbol{q},\Omega)=
 \frac{g \gamma^2}{2A \hbar^2 v_x^2} \sum_{\bsk}    \bigg\{ &&
 \Theta(\mu- k- k\eta \cos\theta_k) \bigg(\frac{1+\cos(\theta_k-\theta_{k'})}{\Omega+ k-k'+i \ep} +  \frac{1-\cos(\theta_k-\theta_{k'})}{\Omega+ k+k'+i \ep}\bigg)
- \nn\\&&\Theta(\mu- k+k\eta \cos\theta_k)
\bigg(\frac{1+\cos(\theta_k-\theta_{k'})}{\Omega- k+k'+i \ep} +  \frac{1-\cos(\theta_k-\theta_{k'})}{\Omega- k-k'+i \ep}\bigg)
\bigg\}.
 \label{pol d app}
 \eearr
 It consists of two parts  that can be transformed to each other by $\Omega\rightarrow-\Omega$ and $\eta\rightarrow-\eta$. 
 The important point in doing the rest of calculation is that in Eq.~\eqref{paid} the step function depends on the direction of wave vector $\bsk$ with respect to $x$ axis.
In what follows we change the integration variable $\theta_k$ as, $\theta+\phi\to \theta$ which causes the denominators to be independent to the direction $\phi$ 
of $\bsq$ and we will have $\bsk'=\sqrt{k^2+q^2+2 k q \cos\theta}$. But then step functions corresponding to Fermi occupation numbers will depend on 
the direction of both $\bsq$ and $\bsk$. This makes the angular integration slightly more complicated than the cases without angular-dependent energy spectrum -- like graphene. 
It turns out to be more convenient if we first perform the $k$ integration and finally do the angular integration on $\theta$ by using the calculus of residues~\cite{Arfken}. 

\subsection{Real part}
The real part of polarization $\Delta\chi$ can be represented as a Cauchy principal value. 
This amounts to setting the imaginary part in the denominator of Eq.~\eqref{paid} equal to zero, 
 \be
\Re\Delta \chi(\boldsymbol{q},\Omega)=
 \frac{g \gamma^2}{4 \pi^2 \hbar^2 v_x^2} \int d\bsk   \bigg\{ 
 \Theta(\mu- k- k\eta \cos(\theta+\phi)) \frac{(\Omega+ k)+ k' m}{(\Omega+ k)^2-(k')^2}
+\Theta(\mu- k+k\eta \cos(\theta+\phi))
 \frac{(-\Omega+k)+ k' m}{(-\Omega+ k)^2-(k')^2}\bigg\},\nn
 \label{paidappendix}
 \ee
where $m=\cos(\theta_k-\theta_{k'})$. 
Now we do integration on $k$, where the step function determines the upper limit $k$ as,
 \be
 \Re\Delta \chi(\boldsymbol{q},\Omega)=
  \frac{g \gamma^2}{4 \pi^2 \hbar^2 v_x^2}  \int d\theta  \int_0^{\mu/(1+\eta \cos(\theta+\phi))} k dk   
 \frac{(\Omega+ k)+ k' m}{(\Omega+ k)^2-(k')^2}
 +\int d\theta  \int_0^{\mu/(1-\eta \cos(\theta+\phi))} k dk
  \frac{(-\Omega+k)+ k' m}{(-\Omega+ k)^2-(k')^2}.\nn
  \label{paidintk}
  \ee
 In the following, we separate first ($+\Omega$) and second ($-\Omega$) term of above relation. The integration on $k$ gives combination of simple 
fractions of and logarithmic terms. The important technical point to notice is that, since the log terms appear in definite integral which involves the
difference between the log functions at two integration limits, it will be meaningful when (i) the argument of log is positive definite, or (ii) the
argument of log is negative definite. In the later case, an overall phase of $\pi$ from the two ends cancel out. 
Requiring the argument of log to be positive definite, {\rm or} negative definite  for $0\leq\theta\leq 2\pi$ gives some constraints for
integration on $k$ of each term of above equation. Let us denote the first (second) term of the above integral as $R^+$ ($R^-$). Then we can write, 
 \bearr
 R^+(\boldsymbol{q},\Omega)&&= \frac{g \gamma^2}{4 \pi^2 \hbar^2 v_x^2}  \int d\theta  \int_0^{\mu/(1+\eta \cos(\theta+\phi))} k dk   
  \frac{(\Omega+ k)+ k' m}{(\Omega+ k)^2-(k')^2}\nn\\ &&=\frac{g \gamma^2 }{4\pi^2 \hbar^2 v_x^2} \int d\theta ~ \bigg\{  \frac{\mu (\Omega+q \cos\theta)}{2(1+\eta \cos(\theta+\phi))(\Omega-q \cos\theta)}
  + \frac{\mu^2 }{2(1+\eta \cos(\theta+\phi))^2(\Omega-q \cos\theta)}\nn\\&&- \frac{\mu(\Omega^2-q^2)}{2(1+\eta \cos(\theta+\phi))(\Omega-q \cos\theta)^2} -(\Omega^2-q^2)\bigg(\frac{(\Omega+q\cos\theta) }{4(\Omega-q \cos\theta)^2}-\frac{ (\Omega^2-q^2)}{4(\Omega-q \cos\theta)^3} \bigg) \ln[ K^+(\mu^+)]
    \theta(q-\Omega)\theta(\mu-\mu^+)
   \nn\\&&-(\Omega^2-q^2)\bigg(\frac{(\Omega+q \cos\theta) }{4(\Omega-q \cos\theta)^2}-\frac{ (\Omega^2-q^2)}{4(\Omega-q \cos\theta)^3} \bigg)  \ln[ K^+(\mu)]
   \bigg(\theta(\Omega-q)+\theta(q-\Omega)\theta(\mu^+-\mu)
    \bigg)\bigg\},
   \label{rp}
  \eearr
and
 \bearr
 R^-(\boldsymbol{q},\Omega)&&= \frac{g \gamma^2}{4 \pi^2 \hbar^2 v_x^2}  \int d\theta  \int_0^{\mu/(1-\eta \cos(\theta+\phi))} k dk   
  \frac{(-\Omega+ k)+ k' m}{(-\Omega+ k)^2-(k')^2}\nn\\ &&=\frac{g \gamma^2 }{4\pi^2 \hbar^2 v_x^2} \int d\theta ~ \bigg\{  \frac{\mu (\Omega-q \cos\theta)}{2(1-\eta \cos(\theta+\phi))(\Omega+q \cos\theta)}
  - \frac{\mu^2 }{2(1-\eta \cos(\theta+\phi))^2(\Omega+q \cos\theta)}\nn\\&&+ \frac{\mu(\Omega^2-q^2)}{2(1-\eta \cos(\theta+\phi))(\Omega+q \cos\theta)^2} +(\Omega^2-q^2)\bigg(\frac{(\Omega-q\cos\theta) }{4(\Omega+q \cos\theta)^2}-\frac{ (\Omega^2-q^2)}{4(\Omega+q \cos\theta)^3} \bigg) \ln[ K^-(\mu_1^-)]
    \theta(\Omega-q)\theta(\mu-\mu^-_1)
   \nn\\&&+(\Omega^2-q^2)\bigg(\frac{(\Omega-q\cos\theta) }{4(\Omega+q \cos\theta)^2}-\frac{ (\Omega^2-q^2)}{4(\Omega+q \cos\theta)^3} \bigg) \ln[ K^-(\mu_2^-)]
       \theta(q-\Omega)\theta(\mu-\mu^-_2)
      \nn\\&&+(\Omega^2-q^2)\bigg(\frac{(\Omega-q \cos\theta) }{4(\Omega+q \cos\theta)^2}-\frac{ (\Omega^2-q^2)}{4(\Omega+q \cos\theta)^3} \bigg)  \ln[ K^-(\mu)]
   \bigg(\theta(\Omega-q)\theta(\mu^-_1-\mu)+\theta(q-\Omega)\theta(\mu^-_2-\mu) \bigg)
   \nn\\&&
   +(\Omega^2-q^2)\bigg(\frac{(\Omega-q\cos\theta) }{4(\Omega+q \cos\theta)^2}-\frac{ (\Omega^2-q^2)}{4(\Omega+q \cos\theta)^3} \bigg) (\ln[ K^-(\mu)]- \ln[ K^-(\mu^{'-})])
       \theta(\Omega-q)\theta(\mu-\mu^{'-})
   \bigg\},
   \label{rn}
  \eearr
where 
\be
 K^+(\mu)=1+\frac{2\mu(\Omega-q \cos\theta)}{(1+\eta \cos(\theta+\phi))(\Omega^2-q^2)}, ~~~~~~ 
 K^-(\mu)=1-\frac{2\mu(\Omega+q \cos\theta)}{(1-\eta \cos(\theta+\phi))(\Omega^2-q^2)}.
 \label{kpm.eqn}
\ee
When the $k$ integration runs up to the upper limit given in terms of $\mu$, the argument of the log function given in
Eq.~\eqref{kpm.eqn} might change sign. We need to cut the integral off once the sign change occurs. 
The sign changes from positive (negative) to negative (positive) happen at $\mu^+,\mu^-_1,\mu^-_2$ ($\mu^{'-}$), where
 \be
 \mu^+=(q^2-\Omega^2)\frac{1+\eta \cos(\phi+\upsilon)}{2(\Omega-q \cos\upsilon)},
 ~~~~~~
 \upsilon=\arccos\bigg[\frac{\alpha \beta- \sqrt{\alpha^2 \beta^2-(\beta^2+\zeta^2)(\alpha^2-\zeta^2)} }{(\zeta^2+\beta^2)}\bigg],
\ee
\be
 \mu^-_1=(\Omega^2-q^2)\frac{1-\eta \cos(\phi+\upsilon)}{2(\Omega+q \cos\upsilon)},~~~~~~
  \upsilon=-\arccos\bigg[\frac{-\alpha \beta+ \sqrt{\alpha^2 \beta^2-(\beta^2+\zeta^2)(\alpha^2-\zeta^2)} }{(\zeta^2+\beta^2)} \bigg],
 \ee
 \be
 \mu^-_2=(\Omega^2-q^2)\frac{1-\eta \cos(\phi+\upsilon)}{2(\Omega+q \cos\upsilon)},~~~~~~
 \upsilon=-\arccos\bigg[\frac{-\alpha \beta- \sqrt{\alpha^2 \beta^2-(\beta^2+\zeta^2)(\alpha^2-\zeta^2)} }{(\zeta^2+\beta^2)} \bigg],
 \ee
\be
 \mu^{'-}=(\Omega^2-q^2)\frac{1-\eta \cos(\phi+\upsilon)}{2(\Omega+q \cos\upsilon)},~~~~~~
 \upsilon=\arccos\bigg[\frac{-\alpha \beta- \sqrt{\alpha^2 \beta^2-(\beta^2+\zeta^2)(\alpha^2-\zeta^2)} }{(\zeta^2+\beta^2)} \bigg].
 \label{nlog}
 \ee
Here in the right hand side of above relations $ \zeta=q+\Omega \eta \cos\phi$ and $\beta= \Omega \eta \sin\phi$ and $\alpha= q \eta \sin\phi$.
The definitions of $\mu^+, \mu^-_1,\mu^-_2$ ($\mu^{'-}$) are (is) such that the arguments $K^\pm$ of logarithm are always positive (negative)
for every value of $0<\theta<2\pi$.

So far we have done the integration on $k$. The next step is to do the angular integration. This can be basically done with the calculus of residues.  
This is straightforward for terms involving fractions of polynomials of trigonometric functions of $\theta$.
When we face the logarithmic function, one has to handle a branch cut. For this, the trick we use is to represent the logarithmic functions in terms of and
integration over some auxiliary variable, $\tau$ as,
\be
\ln K=\int_0^1 \frac{d\tau}{1+\tau K},
\label{ln to int}
\ee
We use this representation and do the angular integration with the calculus of residues. At the end, 
we perform the integration on $\tau$. We can summarize the final result for $R^\pm$
as a piecewise continuous function of the following form,
\bearr
 R^+(\boldsymbol{q},\Omega)=F^0(\bsq,\Omega)+ F^1(\bsq,\Omega) \begin{cases}
       \Theta(\Omega-q) , & G(X^+(\mu))|_0^1 \\
        \Theta(q-\Omega) \Theta(\mu^+-\mu), & G(X^+(\mu))|_0^{x^{+}_u(\mu)}\\
          \Theta(q-\Omega) \Theta(\mu-\mu^+), & G(X^+(\mu^+))|_0^{x^{+}_u(\mu^+)}
     \end{cases},
\label{Rp condition}
\eearr
and
\bearr
 R^-(\boldsymbol{q},\Omega)=F^0(\bsq,\Omega)+ F^1(\bsq,\Omega) \begin{cases}
       \Theta(\Omega-q)  \Theta(\mu^-_1-\mu)  , & G(X^-(\mu))|_0^{x^{-}_u(\mu)} \\
       \Theta(q-\Omega) \Theta(\mu^-_2-\mu) ) , & G(X^-(\mu))|_0^{x^{-}_u(\mu)} \\
       \Theta(\Omega-q)  \Theta(\mu-\mu^-_1) , & G(X^-(\mu_-^1))|_0^{x^{-}_u(\mu^-_1)} \\
        \Theta(q-\Omega) \Theta(\mu-\mu^-_2), & G(X^-(\mu_-^2))|_0^{x^{-}_u(\mu^-_2)}\\
          \Theta(\Omega-q) \Theta(\mu-\mu^{'-}), & G(X^-(\mu_-'))|_0^{x^{-}_u(\mu^{'-})}+\\&G(X^-(\mu))|_0^{x^{-}_u(\mu)}-G(X^-(\mu))|_{x^{-}_d(\mu)}^1
     \end{cases}.
\label{Rn condition}
\eearr
Using the above $R^\pm$, we can summarize Re$\Delta\chi$ as,
\bearr
\Re\Delta\chi(\bsq,\Omega)=F^0(\bsq,\Omega)+ F^1(\bsq,\Omega) \bigg(G(X^+)|_{x_d^+}^{x_u^+} +\sum G(X^-)|_{x_d^-}^{x_u^-}\bigg),
\label{pol doped appe}
\eearr
where the $\pm$ in $X^\pm$ points to the $R^\pm$. The summation $\sum$ in $G(X^-)$ indicates that in the last piece of
Eq.~\eqref{Rn condition} we have three different regions contributing to the integral. 
The functions $F^0,F^1$ and $G$ are given by,
\bearr
&&F^0(\bsq,\Omega)=\frac{g}{16\pi \hbar^2 v_x v_y } \frac{\mu ~q^2}{ \sqrt{1-\eta^2}} \frac{ A(\bsq,\Omega)}{ D^2(\bsq,\Omega)},
\\
&&F^1(\bsq,\Omega)=\frac{g}{16\pi \hbar^2 v_x v_y } \frac{q^2}{  \sqrt{|\Omega^2-q^2|}},
\\
 &&G(x)=B(\bsq,\Omega) x \sqrt{x^2-x'}-\sgn(\Omega-q) \cosh^{-1}\frac{x}{x'},\nn\\&&
\eearr
where $\sgn(\Omega-q)$ is the sign function. The coefficient $A(\bsq,\Omega)$, $B(\bsq,\Omega)$, $D(\bsq,\Omega)$
and $X^{\lambda=\pm}$ have the following definitions,
 \bearr
 &&A(\bsq,\Omega)=q^2(\eta^4+8 \eta^2-8)+4\eta^4 \Omega^24 q \eta \Omega 
  \cos\phi (5 \eta^2-4)+4 \eta^2 \Omega^2 (\eta^2-1) \cos2\phi-q \eta^3(4\Omega \cos3\phi+q\eta \cos4\phi),
\nn\\
&& B(\bsq,\Omega)=\big((q+\eta \Omega \cos\phi)^2-\eta^2 (\Omega^2-q^2)\sin^2\phi\big)/D^2(\bsq,\Omega),
~~~
 D(q,\Omega)=(q+\eta \Omega \cos\phi)^2+\eta^2 (\Omega^2-q^2)\sin^2\phi,
\eearr
\be
X^{\lambda}=2\tilde{\mu} x+ \lambda (\Omega+q\eta \cos\phi),~~~~~~~~~~~~ x'=(\Omega+q \eta \cos\phi)^2-(1-\eta^2)(\Omega^2-q^2),
\ee
\be
x^{-}_{u(d)}=\frac{1}{2 \mu}\bigg( (\Omega+q \eta \cos\phi)-(+)\sqrt{ x'}\bigg),~~~~~~~~~~x_{u}^{+}=-\frac{1}{2 \mu}\bigg( (\Omega+q \eta \cos\phi)+\sqrt{ x'}\bigg)
\ee
Here $\tilde{\mu}$ represents either of $\mu^+,\mu^-_1,\mu^-_2,\mu^{'-}$, depending on which piece of the
$R^\pm$ functions in Eqs.~\eqref{Rp condition} and~\eqref{Rn condition} supports the value of Re$\Delta\chi$. 
The abover results are benchmarked in Fig.~\ref{suzumura.fig} against the numerical results of Ref.~\onlinecite{Nishine2010}.

\subsection{Imaginary part}
In order to calculate the imaginary part, we start from Eq.~\eqref{pol d app} and use  $\Im[1/(x+i\ep)]=-\pi\delta(x)$ 
to write,
\bearr
\Im\Delta \chi(\boldsymbol{q},\Omega)=&&
 -\frac{g \pi \gamma^2}{2A \hbar^2 v_x^2} \sum_{\bsk}   
 \Theta(\mu- k- k\eta \cos\theta_k) \bigg((1+\cos(\theta_k-\theta_{k'}))\delta(\Omega+ k-k') +  (1-\cos(\theta_k-\theta_{k'}))\delta(\Omega+ k+k')\bigg)
\nn\\&&-\Theta(\mu- k+k\eta \cos\theta_k)
\bigg((1+\cos(\theta_k-\theta_{k'}))\delta(-\Omega+k-k') +  (1-\cos(\theta_k-\theta_{k'}))\delta(-\Omega+ k+k')\bigg).
 \label{impol d app}
 \eearr
 We only need to evaluate the above function for positive $\Omega$ and the negative $\Omega$ results 
 can be obtained by appropriate symmetry relations. This assumption makes the second term in above relation irrelevant
 as the Dirac delta function does not pick any pole.  Therefor we are left with the positive $\Omega$ contribution
 from the first term that splits into three parts,
\bearr
 &&I_1(\bsq,\Omega)= -\frac{g  \gamma^2}{8 \pi \hbar^2 v_x^2} \int d\bsk 
  \theta(\mu- k- k\eta \cos\theta) (1+\cos(\theta_k-\theta_{k'}))\delta(\Omega+ k-k'),
 \nn\\&&  I_2(\bsq,\Omega)=  \frac{g  \gamma^2}{8 \pi \hbar^2 v_x^2} \int d\bsk 
   \theta(\mu- k+k\eta \cos\theta) (1+\cos(\theta_k-\theta_{k'}))\delta(-\Omega+ k-k'),
 \nn\\&&
 I_3(\bsq,\Omega)=  \frac{g  \gamma^2}{8 \pi \hbar^2 v_x^2} \int d\bsk 
    \theta(\mu- k+k\eta \cos\theta) (1+\cos(\theta_k-\theta_{k'}))\delta(-\Omega+ k+k').
\eearr  
By change of variable, $\theta\rightarrow\theta+\phi$,  and using $\delta[f(z)]=\delta(z-z_0)/|f'(z_0)|$, 
we rewrite the delta functions as $\delta(\theta-\theta_0)/|k q \sin\theta_0|$. In this equation, the $\cos\theta_0$ for each delta function 
has its own definition, and obviously $\sin\theta_0=\pm \sqrt{1-\cos^2\theta_0}$. The trigonometric inequality  $-1\leq\cos\theta_0\leq1$ 
places some constraint on the $q$ and $\Omega$ as follows,
\bearr
 \begin{cases}
       \delta(\Omega+ k-k') :~~~~~\cos\theta_0=(\Omega^2+2\Omega k-q^2)/2 k q  , & \Omega+2k>q \\
       \delta(-\Omega+ k-k') :~~~\cos\theta_0=(\Omega^2-2\Omega k-q^2)/2 k q  , & -\Omega+2k>q \\
       \delta(-\Omega+ k+k') :~~~\cos\theta_0=(\Omega^2-2\Omega k-q^2)/2 k q  , & \Omega-q<2k<\Omega+q
 \end{cases}.
\label{delta condition}
\eearr  
With the above expressions for $\cos\theta_0$ and $\sin\theta_0$, the above three integrals can be
evaluated to give,
\bearr
I_1(\bsq,\Omega)=&& -\frac{g  \gamma^2}{32 \pi \hbar^2 v_x^2} \frac{q^2}{\sqrt{q^2-\Omega^2}} \Theta(q-\Omega) \int_1^{\infty} dp \sqrt{p^2-1}  ~\theta (a^+(\bsq,\Omega)-b^+(\bsq,\Omega) p\pm c^+(\bsq,\omega) \sqrt{p^2-1}) \nn\\&&=-\frac{g  \gamma^2}{32 \pi \hbar^2 v_x^2} \frac{q^2}{\sqrt{q^2-\Omega^2}} \theta(q-\Omega) 
   \bigg\{ \Theta(a^+(\bsq,\Omega)+b^+(\bsq,\Omega)) (G_0^+(x)|_1^{r^+_1}+G_0^+(x)|_1^{r^+_2})\nn\\&&+\Theta(-a^+(\bsq,\Omega)-b^+(\bsq,\Omega)) \Theta(a^+(\bsq,\Omega)^2-b^+(\bsq,\Omega)^2+c^+(\bsq,\Omega)^2) G_0^+(x)|_{r^+_1}^{r^+_2}
   \bigg\},
\label{im1.eqn}
\eearr
\bearr
I_2(\bsq,\Omega)=&& \frac{g  \gamma^2}{32 \pi \hbar^2 v_x^2} \frac{q^2}{\sqrt{q^2-\Omega^2}} \Theta(q-\Omega) \int_1^{\infty} dp \sqrt{p^2-1}  ~\Theta (a^-(\bsq,\Omega)-b^-(\bsq,\Omega) p\pm c^-(\bsq,\omega) \sqrt{p^2-1}) \nn\\&&=\frac{g  \gamma^2}{32 \pi \hbar^2 v_x^2} \frac{q^2}{\sqrt{q^2-\Omega^2}} \Theta(q-\Omega) 
   \bigg\{ \Theta(a^-(\bsq,\Omega)+b^-(\bsq,\Omega)) (G_0^+(x)|_1^{r^-_1}+G_0^+(x)|_1^{r^-_2})\nn\\&&+\Theta(-a^-(\bsq,\Omega)-b^-(\bsq,\Omega)) \Theta(a^-(\bsq,\Omega)^2-b^-(\bsq,\Omega)^2+c^-(\bsq,\Omega)^2) G_0^+(x)|_{r^-_1}^{r^-_2}
   \bigg\},
\label{im2.eqn}
\eearr
\bearr
I_3(\bsq,\Omega)=&& \frac{g  \gamma^2}{32 \pi \hbar^2 v_x^2} \frac{q^2}{\sqrt{\Omega^2-q^2}} \Theta(\Omega-q) \int_{-1}^{1} dp \sqrt{1-p^2}  ~\Theta (a^-(\bsq,\Omega)-b^-(\bsq,\Omega) p\pm c^-(\bsq,\omega) \sqrt{p^2-1}) \nn\\&&=\frac{g  \gamma^2}{32 \pi \hbar^2 v_x^2} \frac{q^2}{\sqrt{\Omega^2-q^2}} \Theta(\Omega-q) 
   \bigg\{  2 \Theta(a^-(\bsq,\Omega)+b^-(\bsq,\Omega)) G_0^-(x)|_{-1}^{1}+\nn\\&&
   \Theta(a^-(\bsq,\Omega)+b^-(\bsq,\Omega)) \Theta(\Delta^-(\bsq,\Omega)) \Theta(1-|u(\bsq,\Omega)|) (G_0^-(x)|_{-1}^{r^-_1}+G_0^-(x)|_{-1}^{r^-_2})
  +\nn\\&&
     \Theta(a^-(\bsq,\Omega)+b^-(\bsq,\Omega)) \Theta(\Delta^-(\bsq,\Omega)) \Theta(-1-u(\bsq,\Omega)) G_0^-(x)|_{r^-_1}^{r^-_2}
   -\nn\\&&\Theta(-a^-(\bsq,\Omega)-b^-(\bsq,\Omega))  \Theta(\Delta^-(\bsq,\Omega)) \Theta(|u(\bsq,\Omega)|-1) G_0^-(x)|_{r^-_1}^{r^-_2}+
   \nn\\&&\Theta(-a^-(\bsq,\Omega)-b^-(\bsq,\Omega))  \Theta(\Delta^-(\bsq,\Omega)) \Theta(1-|u(\bsq,\Omega)|) (G_0^-(x)|_{-1}^{r^-_2}+ G_0^-(x)|_{-1}^{r^-_2})
   \bigg\},
\label{im3.eqn}
\eearr
In above equations, $p=(2k\pm\Omega)/q$, where $+$ stands for $I_1$ and $-$ stands for $I_2 , I_3$. The definition of 
functions used in the above relations is given by,
\be
G_0^+(x)=x\sqrt{x^2-1}-\cosh^{-1}(x),  ~~~~~~~~~~~ G_0^-(x)=x\sqrt{1-x^2}+\sin^{-1}(x),
\ee
\be
r^{1(2)}_{\lambda}=\frac{-a^{\lambda}(\bsq,\Omega) b^{\lambda}(\bsq,\Omega) -(+) c^{\lambda}(\bsq,\Omega)+\lambda \sqrt{\Delta^{\lambda}(\bsq,\Omega)}}{
\left(b^{\lambda}(\bsq,\Omega)\right)^2- \left(c^{\lambda}(\bsq,\Omega)\right)^2 },~~~~
\Delta^{\lambda}(\bsq,\Omega)=\left(a^{\lambda}(\bsq,\Omega)\right)^2-\left(b^{\lambda}(\bsq,\Omega)\right)^2+\left(c^{\lambda}(\bsq,\Omega)\right)^2,
\ee
where $\lambda=\pm$ and
\bearr
&& a^{\pm}(\bsq,\Omega)=2\mu \pm \Omega\pm q \eta \cos\phi, ~~~~
b^{\lambda}(\bsq,\Omega)=-(q+\Omega \eta \cos\phi ),\nn\\
&&c^{\lambda}(\bsq,\Omega)=\pm\eta \sin\phi \sqrt{|q^2-\Omega^2|},~~~
u(\bsq,\Omega)=\frac{a^-(\bsq,\Omega)}{b(\bsq,\Omega)}.
\eearr

\end{widetext} 

\bibliography{mybib}

\end{document}